\documentclass[9pt]{spie}

\usepackage{wrapfig}
\usepackage{indentfirst}
\usepackage{graphicx}
\usepackage{amsmath}
\usepackage{hyperref} 
\title{Small-brain neural networks rapidly solve inverse problems with vortex Fourier encoders}
\author{Baurzhan Muminov and Luat T. Vuong$^{*}$
\skiplinehalf
{Department of Mechanical Engineering, University of California at Riverside, Riverside, CA, 92521, USA}\\
\supit{*}{Corresponding author: LuatV@UCR.edu}
}

\begin{document}
\maketitle
\begin{abstract}
We introduce a vortex phase transform with a lenslet-array to accompany shallow, dense, ``small-brain'' neural networks for high-speed and low-light imaging. Our single-shot ptychographic approach exploits the coherent diffraction, compact representation, and edge enhancement of Fourier-tranformed spiral-phase gradients. With vortex spatial encoding, a small brain is trained to deconvolve images at rates 5-20 times faster than those achieved with random encoding schemes, where greater advantages are gained in the presence of noise. Once trained, the small brain reconstructs an object from intensity-only data, solving an inverse mapping without performing iterations on each image and without deep-learning schemes. With this hybrid, optical-digital, vortex Fourier encoded, small-brain scheme, we reconstruct MNIST Fashion objects illuminated with low-light flux (5 nJ/cm$^2$) at a rate of several thousand frames per second on a 15 W central processing unit, two orders of magnitude faster than convolutional neural networks.
\end{abstract}

\section{Introduction}
Our ability to solve inverse problems and reconstruct object features from either incomplete or mixed-signal components is essential for a broad range of applications, from x-ray imaging to remote sensing. Reconstruction or deconvolution of an object pattern from sensor data is often challenging from a practical standpoint, since algorithms must address the famous Phase Problem in which the phase information is lost by the sensor, which only registers photonic magnitude or intensity. Iterative approaches have been developed but are time-consuming, since the process may require multiple restarts with several initial guesses until convergence is achieved \cite{Latychevskaia2018}. It is fascinating how this area{\normalfont ---}iterative solutions of the Phase Problem{\normalfont ---}has developed and given rise to a set of optimization techniques that are today applied in other domains \cite{Elser2007} and notably, provide the capacity to image through turbid and scattering media \cite{Bertolotti2012, Katz2014}. 

Recently, it is also possible to obviate the Phase Problem for image construction with computational imaging. An expanding research area involves the application of neural networks, specifically deep-learning convolutional neural networks (CNNs) \cite{Barbastathis2019}. With CNNs, the recording of an interference pattern such as a hologram, or several overlapping snapshots as with ptychography, can be used to reproduce object features \cite{Tahara2018, Rivenson2019, Konda2020}. When using coherent diffraction through phase masks, at least two distinct images are needed to attempt the Phase Problem \cite{Sidorenko2015, Zhang2013, Wang2020}. The first (to the authors' knowledge) application of CNNs for image reconstruction, is presented in \cite{Sinha2017}, where a phase-encoded image on a spatial light modulator is reconstructed via CNNs using intensity data from the camera.  "Non-line-of-sight" CNN imaging is recently demonstrated from albedo autocorrelation patterns of speckles \cite{Metzler2020, OToole2018, Lei2019}. In these examples, as well as others that leverage neural networks, it is possible to reconstruct an {\it object type} without solving the Phase Problem, i.e., one may successfully predict an object using prior-trained patterns without being able to identify the \textit{position} of the object \cite{Metzler2020}. 

Still, deep learning neural networks offer additional functionality in the process of reconstructing the object. For example, simultaneous autofocusing with phase recovery \cite{Wu2018} or super-resolution in pixel-limited of diffraction-limited systems \cite{Liu2019}. With sets of training and testing diffusers, the phase information encoded through controlled speckle patterns can be leveraged to predict the outputs from previously unseen diffusers \cite{Li2018}. The non-exhaustive list of important applications include profilometry \cite{Song2016}, imaging through smoke \cite{Locatelli2013}, and LIDAR that leverages multiple point cloud and time-of-flight information \cite{Mitchell2018}. Additional examples of ``nonlinear reservoir learning'' are presented in \cite{Antipa2017}, which employs caustic patterns for original object reconstruction. The challenge with deep learning methods, however, is that the neural network requires large training sets, long training times, and these neural networks have higher degrees of computational complexity that render them vulnerable to adversarial network attacks \cite{Goodfellow2018}. 
  
In this paper, we focus on the application of shallow and dense neural networks and ask, is it possible to achieve additional image-reconstruction functions without deep-learning and without iterative schemes? Such ``small brain'' approaches are regression-based and provide the advantage of a single forward pass, i.e., no requisite iterative phase retrieval procedures \cite{Vuong2019, Horisaki2016}. We demonstrate a new approach to image reconstruction with optical preprocessing in lieu of CNNs. Our strategy is similar to other hybrid and diffractive optical neural network approaches that aim to offload mathematical computation to the propagation of light \cite{Zhou2020, Khoram2019, Cordaro2019, Chang2018, Zhu2017, Guo2018, Lin2018, OToole2018}. What our scheme shows, unlike others, is that a small brain is capable of solving the inverse mapping with vortex spatial encoding in the Fourier domain. Moreover, the inverse mapping is performed efficiently and with less computational complexity with vortices than with random encoded patterns. This indicates that the optical vortex provides feature extraction in the Fourier representation, which further reduces the computational load. This article is an expanded version of work recently presented at \cite{Muminov2020}.

\begin{wrapfigure}{r}{0.5\textwidth}
    \centering
    \includegraphics[width=0.5\textwidth]{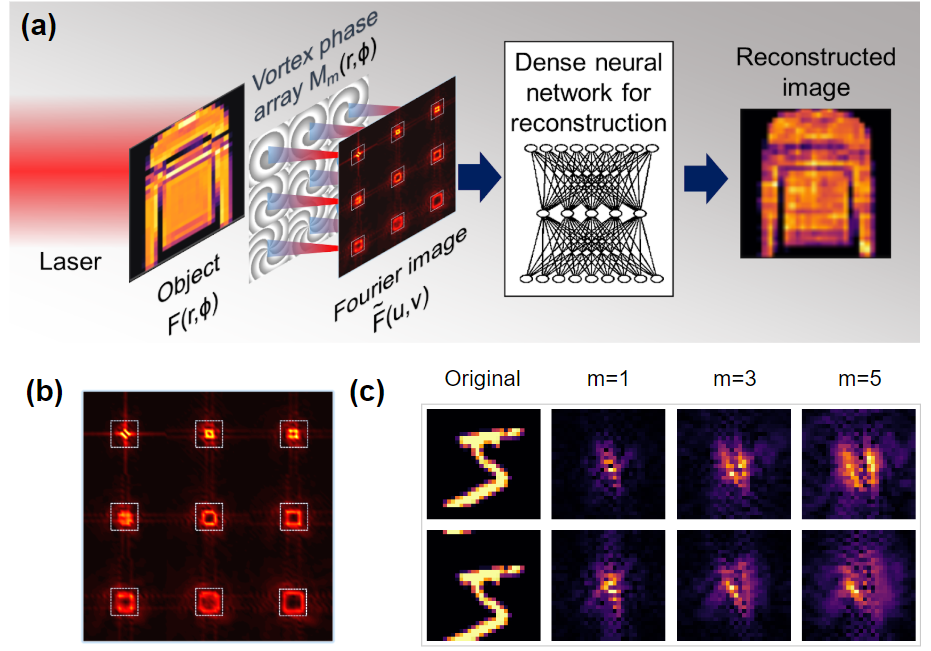}
    \caption{(a) The general schematic of the technique: a laser illuminates the object. Transmitted light is phase-modulated with a multi-vortex lens array. The back focal plane vortex-Fourier intensity patterns are fed to a neural net that reconstructs the original image. (b) The vortex-Fourier patterns have fewer pixels. Here, the combined area of 9 dotted squares is equivalent to the area of the original object. (c) The vortex-Fourier pattern for centered and shifted MNIST handwritten digit '5' show increasing sensitivity to shifts and larger areas (lower intensities) with increasing topological charges $m$.}
    \label{fig:Base_scheme}
\end{wrapfigure} 

Our approach provides new capacity to successfully image under low-light signal conditions. The results are dramatic since Fourier representations are compressed (i.e., the illuminated area of the camera sensor is much smaller than the area illuminated by the object's real image) \cite{Balboa2003} and robust (the resulting computer vision scheme is not susceptible to the rapid variations in scene illumination) \cite{Krahmer2014}.  We note that, from a purely computational standpoint, Fourier representations have been demonstrated to be efficient at solving classification problems \cite{Rippel_spectral, Trainable_Spectrally, Popa2018}. Object reconstruction with Fourier representations reduces requisite memory, power, or energy requirements and may even achieve real-time image processing \cite{Chen2016cnn, Cherukara2018}. The advantages of Fourier operations further multiply since they may completed optically before the digital neural network \cite{Chang2018}. Still, in each of the aforementioned cases using deep learning, the transferability of learned maps remains an issue{\normalfont ---}i.e., the trained neural nets are task-specific and, moreover, equipment-specific. This issue of transferability is further addressed in our work. We demonstrate that our small-brain approach does not require specificity in the trained data to solve the inverse problem. In fact, overcoming the Phase Problem with low computational complexity is our milestone result.

To achieve this, we exploit topological representations with optical vortices{\normalfont ---}more specifically, with Laguerre-Gaussian beams. Such beams with spiral phase gradients are characterized by a topological charge and associated with phase singularities at which the electric field is strictly zero \cite{Wang2018, Allen1992, Soskin1997, Lee2006, Dennis2009, Zhu2019,  Chen2016}. A famous example that leverages phase singularities for imaging is the ``vortex coronograph,'' in which a vortex phase is placed in the Fourier imaging plane. A higher-resolution vortex camera is recently demonstrated in \cite{Novak2020} where the reconstruction contrast ratio is increased as a result of the vortex phase. 

In our approach, we achieve non-iterative, single-shot object reconstruction with a topological vortex-based lenslet-array design that contains multiple vortex phases in a lenslet pattern using the resulting edge-enhanced Fourier-plane representations. The presence of the vortex provides spatial encoding to break the translation invariance of the measured Fourier pattern and solve the Phase Problem. Image reconstruction is performed with dense neural nets or shallow neural nets. Again, we refer to this few-hidden-layer neural network that does not use deep learning as a "small-brain" \cite{Vuong2019}.

\section{innovation}
Figure \ref{fig:Base_scheme} depicts our imaging scheme, where multiple images of the object $F(r,\phi)$ are collected in the Fourier domain: the light transmitted through each lenslet is modulated by different vortex and lens mask patterns $M_m(r,\phi)$; and the camera detects the scaled, modulus-squared image of the Fresnel-propagated, vortex-Fourier-transformed intensity patterns, $|\tilde{F}_{m}(u,v)|^2$. Here, $m$ is the vortex topological charge, $r$ and $\phi$ are the real domain cylindrical coordinates, and $u$ and $v$ are the Fourier-plane Cartesian coordinates. The vortex Fourier intensity patterns $\tilde{F}$ are concentrated in a relatively small area but are typically donut-shaped with a wider donut as $m$ increases \ref{fig:Base_scheme}(b)]. The vortex phase in the object `real-domain' spatially encodes and breaks translational invariance of the Fourier-transformed intensity pattern \ref{fig:Base_scheme}(c)]. 

We consider a few small-image datasets as object inputs and compare different representations in $F(r,\phi)$. For each positive, real-valued dataset image $\mathbf{X}$, we map the phase changes: 
\begin{equation}
F(r,\phi) = e^{i\alpha_0 \mathbf{X}} \label{Feqphase}
\end{equation}
where $\alpha_0$ is the dynamic range of the object phase-shift. This mapping is convenient because the signal power is invariant with our choice of $\mathbf{X}$. We have also considered opaque objects where $\mathbf{X}$ blocks or absorbs the signal, i.e., $F(u,v)\propto \mathbf{X}$, which yields similar trends. 

There are three primary innovations in our results. We demonstrate: 1) edge enhancement of spectral features with a vortex lens; 2) rapid inverse reconstruction of the image without a similar, learned dataset; and 3) robustness to noise, which depends on the layer activations. 

\subsection{Vortex Fourier encoding and feature extraction}
Here, we make two new claims about the special spatial qualities of optical vortices for image processing, namely:
\begin{itemize}
    \item edge enhancement of the Fourier pattern, and
    \item ptychographic mixing of real and imaginary parts to preserve phase in the intensity measurements.
\end{itemize} 
Consider a Fourier-space representation of a simple lenslet pattern composed of multiple-$m$ vortex phases,
\begin{equation}
M_m(r,\phi)=\begin{cases} 
      e^{\frac{-i\pi r^2}{\lambda f}+i m \phi} & 0\leq r< a \\
      0 & otherwise\\
   \end{cases} \label{Meq}
\end{equation}
where $a$ is the radius of the mask aperture, $\lambda$ is the wavelength and $f$ is an effective focal length. This pattern with a centered vortex is appropriate for our dataset's mostly-centered image objects $\mathbf{X}$. Our image reconstruction approach does not require that $m$ is an integer. 
At the phase plate [Eq. \ref{Meq}], the transmitted pattern is a sum of Laguerre-Gaussian modes at $z=0$ with different radial indices $p$, 
\begin{equation}
M_m(r,\phi) = \sum_p W_{p}LG_{m, p}(r,\phi) \label{LGmodal},
\end{equation}
where $W_{p}$ are modal coefficients related to index $p$ and associated with Laguerre-Gaussian profiles, which we separate into components,
\begin{eqnarray}
    LG_{m, p}(r, \phi) &=& L_p^{|m|}(2r^2/w^2) R(r) G(r) V_m(r, \phi) \\
    R (r) &=&  e^{\frac{-i\pi r^2}{\lambda f}} \\
    G (r)&=&  \frac{1}{w}e^{-(\frac{r}{w})^2}\\
    V_{m}(r, \phi) &=& A_{m,p}r^{|m|} e^{i m \phi},
\end{eqnarray} 
where $L_p^{|m|}(2r^2/w^2)$ are the generalized Laguerre polynomials that depend on $m$ and $p$, and $A_{m,p} = \sqrt{\frac{2^{|m|+1}p!}{\pi(p+|m|)!}}w^{-|m|}$ \cite{Plick2015}. We expect that the waist of the beam $w$ is larger than the features of the object $F$. The modal coefficients are,
\begin{equation}
W_p =\int \int M_m LG^*_{m, p}(r, \phi) rdrd\phi= A_{m,p} \int_0^a 2\pi r^{|m|+1}L_p^{|m|}(2r^2/w^2)G (r)dr.
\end{equation} 
The phase-singular term $V_m$ is a radial magnitude gradient and azimuthal phase gradient, which can be simplified \cite{Rasouli2019}  
\begin{eqnarray}
    V_{m}(r, \phi)  = A_{m,p} r^{|m|} e^{i m \phi}
    = A_{m,p}[r\cos(\phi)+i \text{sgn}(m)\sin(\phi)]^{|m|} \label{Veq}.
\end{eqnarray}
Since $r\cos(\phi)$ and $r\sin(\phi)$ are the canonical $x$ and $y$ Cartesian coordinates, which are Fourier-transform pairs with $\frac{u}{f \lambda}$ and $\frac{v}{f \lambda}$  \cite{goodman1996},
\begin{equation}
    \tilde{V}_m(u,v) = \mathcal{F} \{ V_{m}(r, \phi) \} 
    = A_{m,p}\bigg{(}\lambda f\bigg{[}\text{sgn}(m) \frac{\partial}{\partial v}-i\frac{\partial}{\partial u}\bigg{]}\bigg{)}^{|m|}, \label{Vict}
\end{equation}
where $\mathcal{F}$ is the 2D Fourier transform operator. We view $\tilde{V}_m$ as a linear differential operator for the inputs to our neural network, which are the intensity patterns in the back focal plane \cite{goodman1996},
\begin{eqnarray}
 \mathbf{Y} &=&|\tilde{F}_{m}(u,v)|^2 + noise\\
 &=& \bigg{|}\sum_p \frac{W_p}{\lambda f} A_{m,p}\bigg{(}\lambda f\bigg{[}\text{sgn}(m) \frac{\partial}{\partial v}-i\frac{\partial}{\partial u}\bigg{]}\bigg{)}^{|m|} \mathcal{F}\{F(r,\phi)L_p^{|m|}(2r^2/w^2) G (r)\}\bigg{|}^2 + noise\label{FMFeq}.
\end{eqnarray}
In other words, the placement of the optical vortex provides unique optical preprocessing for the Fourier-plane data. Notably, the detected intensity patterns [Eq. \ref{FMFeq}] are composed of real and imaginary differentials of the Fourier transform of $F$. This differential scheme provides feature extraction via differentials in a manner similar to that deployed in the HERALDO method for image reconstruction \cite{HERALDO}. At the same time, the vortex phase mixes the real and imaginary field components, to produce fringes in the intensity pattern. 

We emphasize that the chirality in the system is critical. $V_m$ represents a radial change in magnitude multiplied by an azimuthal change in phase. The resulting fringes exhibit a specific handedness associated with $m$, which preserves phase in the intensity patterns. For example, while the momentum-space representations of radial Laguerre polynomials \cite{Plick2015} also provide spatial derivatives of Fourier components, they do not provide ptychographic encoding to solve the Phase Problem. 

\subsection{Small-brain machine learning of the inverse reconstruction mapping} \label{inverseR}
Since a neural network is capable of guessing the reconstruction based on pre-learned patterns without solving the inverse or Phase Problem, we take a new approach towards understanding how the neural network learns these patterns. We test the inverse reconstruction of the neural network with several categorically-patterned datasets, namely the Fashion-MNIST \cite{fashion}, Kuzushiji-MNIST \cite{kanji} Arabic\cite{arabic}, as well as the canonical handwritten MNIST digit dataset \cite{digits}. The ``ground truth'' outputs $\mathbf{X}$ are the dataset's 28x28-pixel images and are unit-normalized to provide comparable peak signal to noise (PSNR) with different image types across datasets. The vortex imaging scheme is capable of being applied as a ``camera''; the vortex-based reconstruction achieves a mapping that is transferable or generalizable in cases when a random encoding scheme does not. 

With our scheme, the inputs $\mathbf{Y}$ are the modulus-squared vortex Fourier-transforms of Gaussian-apertured $FG$ [Eq. \ref{FMFeq}]. We set $\alpha_0 = \pi/2$ in a phase modulation scheme [Eq. \ref{Feqphase}]. We set $f\lambda = 0.1$. If there is more than one vortex-Fourier pattern used for reconstruction, the procedure is repeated and the vortex images are catenated and/or truncated for the neural network input $\mathbf{Y}$. A dense, shallow, small-brain neural net with 1 hidden layer is trained with a mean-squared error (MSE) loss function. During ``training'', the neural network is provided a subset of the related $\mathbf{X}$ and $\mathbf{Y}$ and during ``testing'', the neural network is provided $\mathbf{Y}$ to solve for $\mathbf{X}$. The testing image set that the neural net has not seen before is referred to as the ``validation'' set.

\begin{wrapfigure}{r}{0.5\textwidth}
  \begin{centering}
    \includegraphics[width=0.5\textwidth]{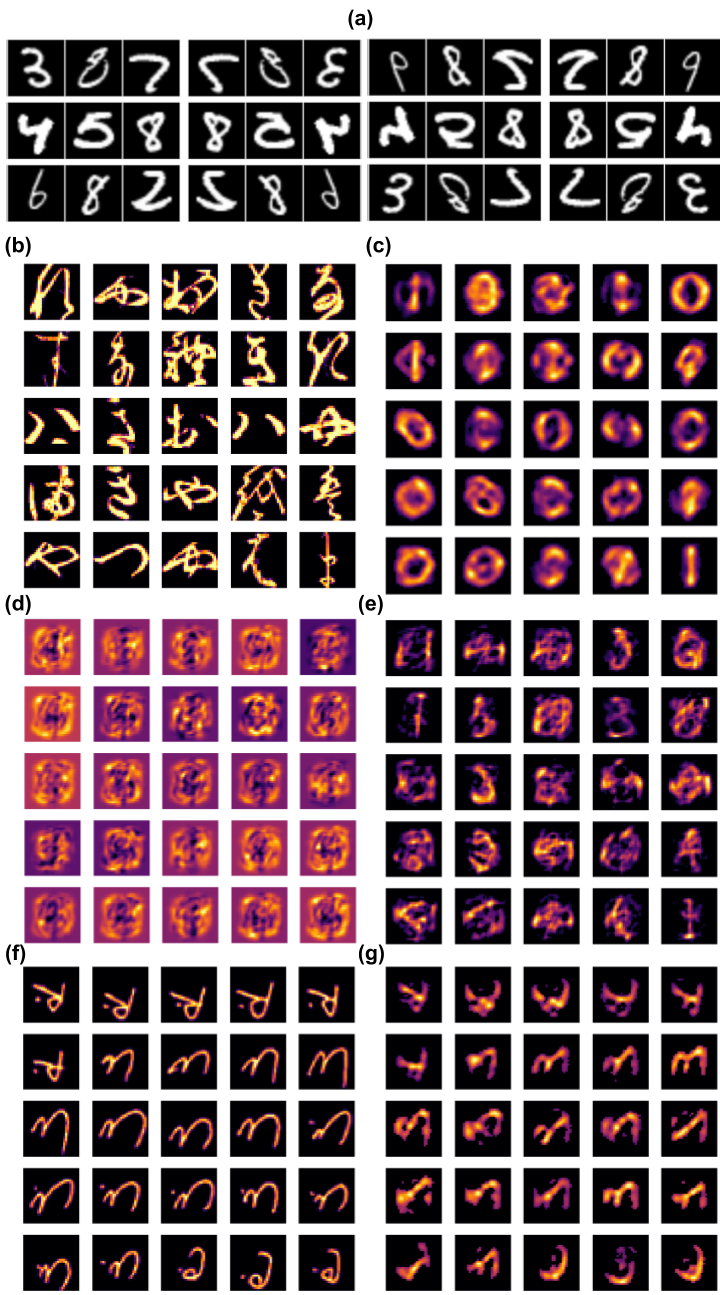}
    \end{centering}
    \caption{(a) 36 images of the training set composed of MNIST handwritten digit images flipped vertically and horizontally. (b) 25 images of the test set of MNIST Kuzushiji. Reconstructed images with (c) no spatial encoding, (d) random spatial encoding, and (e) vortex spatial encoding with $m=3$. When we train with the same MNIST digits in (a) and (f) test with Arabic letters, (g) the reconstructed images with vortex Fourier encoding and $m=3$ are in good agreement. With vortex Fourier encoding, the neural network produces a generalizable inverse map for image reconstruction.}
    \label{fig:Kanji_main}
\end{wrapfigure}

The importance of spatial encoding in reconstruction is shown in Fig. \ref{fig:Kanji_main}. We train the simulations with numbers from the MNIST hand-written number dataset (flipped and inverted) [Fig. \ref{fig:Kanji_main}(a)]. However, we test with a separate dataset with patterns that are unrelated and different. With Kuzushiji characters [Fig. \ref{fig:Kanji_main}(b)], the reconstruction fails when there is no spatial encoding and also fails with a random spatial encoder [Fig. \ref{fig:Kanji_main}(c-d)], but succeeds, with some loss of resolution, with a vortex Fourier encoder ($m=3$) [Fig. \ref{fig:Kanji_main}(e)]. Without both compressed and encoded inputs and without previous patterns for guessing, the neural network cannot produce an inverse map from $\mathbf{Y}$ to $\mathbf{X}$. With the Arabic data set [Fig \ref{fig:Kanji_main}(f-g)] the reconstructed letters are impressive considering that we limit our training to the types of handwritten digits that deviate substantially from the formal Arabic letters. This illustration shows one approach to testing our intuition about the Phase Problem with neural networks and also demonstrates our unique opportunities with vortex-Fourier encoding schemes: a combination of compressed, encoded inputs is critical.

\begin{figure}
    \centering
    \includegraphics[width=0.9\textwidth]{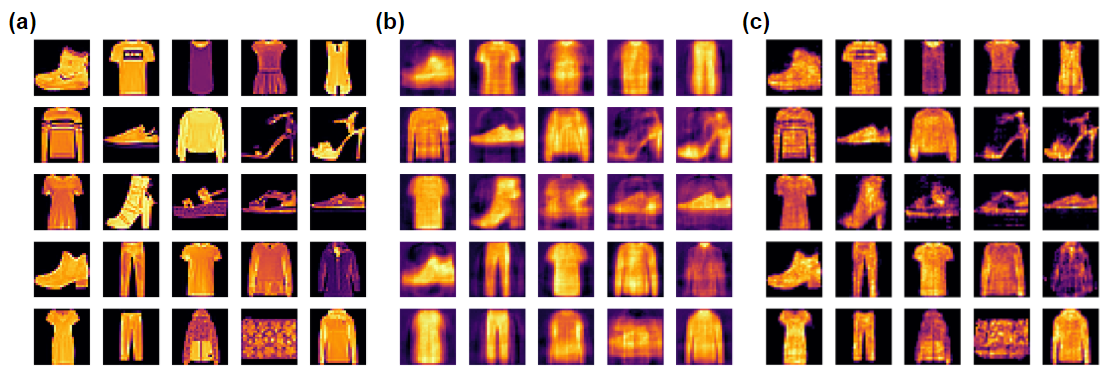}
    \caption{(a) 25 of the original Fashion-MNIST dataset images. Reconstruction with (b) $m=0$, with no spatial encoding and (c) two vortices of topological charges $m=1,3$ with non-linear activation is used in the last layer. Without spatial encoding, the neural network still learns the patterns when there are categorical variations between training and test sets. The structural similarity index metric (SSIM) is quantified in Table 1. }
    \label{fig:rec_result_theor}
\end{figure}

We repeat the machine-learning problem with the Fashion MNIST dataset, where the training and testing sets are more similar, and where, subsequently, the neural network is able to provide the inverse mapping of the $\mathbf{Y}$ to $\mathbf{X}$ without spatial encoding. Fig. \ref{fig:rec_result_theor}(a) shows the validation set. Even though the neural network has not seen the validation set before, unlike in the previous example, it has been trained with similar sets of images that fall into various categories (shirts, shoes, dresses, etc.,). Figure \ref{fig:rec_result_theor}(b) shows discernible reconstructions of images without spatial encoding. In this case, the neural network has learned and reconstructed patterns. The reconstructed images exhibit ghosting as a result of this uncertainty. Again, it is important to emphasize that even though the neural network is able to reconstruct the MNIST images, it does so with learned similarities with the training set, which is not an inverse mapping. 

With a vortex pattern, the Fourier-image phase is preserved and encoded, and the reconstructed images are impressively sharp and delineated. This is an important remark as the ghosting or faded silhouettes are problematic for classification and computer vision algorithms \cite{Baker2018}. Table 1 illustrates the convergence of the reconstructed images of the Fashion MNIST dataset to the original, both in terms of SSIM and MSE, as the number of vortices increases. We also employ the Structural Similarity Index Metrics (SSIM), which is limited to a [0,1] segment, which is a more reasonable metric for human quality perception evaluation \cite{ZhouWang2009}. For comparison, Table 1 also shows the SSIM and MSE for three-layer CNN-trained reconstruction with single and dual vortex datasets. This comparison suggests that our proposed architecture achieves the same quality while yielding much lower computational overhead (more than 3000 FPS for the proposed network and less than 50 FPS for a three-layer CNN with 3x3 kernel). 


\begin{table}
  \begin{center}
    \caption{SSIM and MSE for Fashion-MNIST reconstruction. The table shows that near-optimized quality is achieved with 2 vortices.}
    \label{tab:table1}
    \begin{tabular}{c|c|c|c|c|c|c|r } 
      \textbf{} &\textbf{Plain Fourier} & \textbf{1 vortex, linear} & \textbf{2 vortices, linear}& \textbf{3 vortices} & \textbf{CNN, 1 vortex} & \textbf{CNN, 2 vortices}\\
      \hline
      SSIM & 0.45 & 0.62 & 0.84 & 0.88 & 0.61 & 0.84\\
      MSE & 0.0280 & 0.0242 & 0.0140 & 0.0122 & 0.0235 & 0.0145\\
    \end{tabular}
  \end{center}
\end{table}

\subsection{Speed and robustness to noise}
\begin{wrapfigure}{r}{0.5\textwidth}
    \centering
    \includegraphics[width=0.5\textwidth]{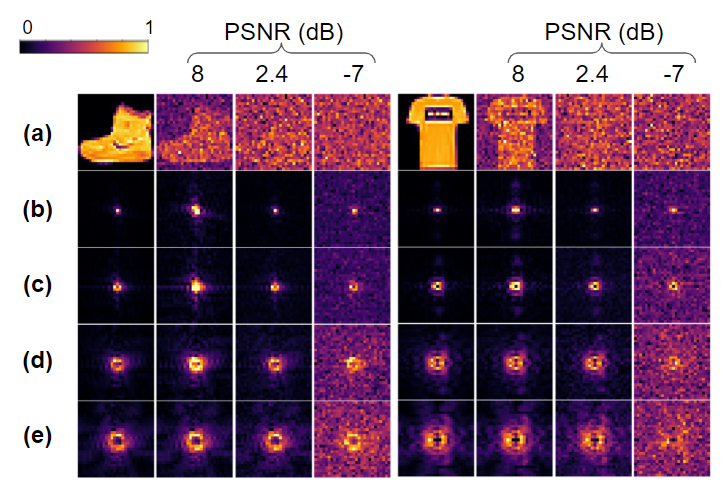}
    \caption{Normalized test inputs $\mathbf{Y}$ to the neural network with noise for two Fashion-MNIST images. Each block shows decreasing peak signal to noise ratio (PSNR) in columns from left to right. (a) Object data $\mathbf{X}$ and vortex spatial encoding (b) $m=0$ or no vortex (c) $m=1$ (d) $m=3$ and (e) $m=5$. The Fourier transform representations have higher PSNR that decreases with higher $m$ given the same camera and light flux for $\mathbf{X}$.}
    \label{NoisyIm}
\end{wrapfigure} 
We study the variable speed and robustness of the reconstruction with vortex Fourier encoding with other random encoding approaches and consider sensor shot and dark noise in the neural network input [Eq. \ref{FMFeq}] \cite{NoiseD}, 
\begin{equation}
    noise = P_n(|\tilde{F}(u,v)|^2) + P_n(\sigma_d^2).
\end{equation}
Both noise terms have Poisson distributions $P_n(\mu)$, where $\mu$ is the expected value and variance. The variance of the sensor noise is proportional to the intensity over the pixel $|\tilde{F}(u,v)|^2$ while the variance of the dark noise $\sigma_d^2$ is related to the dark current and read noise of each pixel. The continuously-valued noise is related to the camera noise with MSE,
\begin{eqnarray}
    MSE &=& \frac{1}{N}\Sigma (y_0-y_i)^2.
\end{eqnarray} 
where $N$ is the number of pixels and $y_0$ and $y_i$ are the noiseless and noisy pixels of $\mathbf{Y}$. 

For our simulations here, we assume that the specifications of the camera do not interfere with the our reconstruction algorithm. Therefore, regardless of the light intensity, we assume that the full dynamic range of a 12-bit camera is used. To vary the noise, we keep $\sigma_d^2$ fixed and change $|\tilde{F}(u,v)|^2$ to study the peak SNR (PSNR) \cite{ZhouWang2009},
\begin{eqnarray}
    PSNR &=& 10\log_{10}\frac{P_{pk,signal}}{P_{noise}}\\
    &=& 10\log_{10}\left(\frac{(2^{L}-1)^2}{MSE}\right),\\
&=&  10\log_{10}\frac{max(|\tilde{F}(u,v)|^2)}{\langle|\tilde{F}(u,v)|^2\rangle + \sigma_d^2}
\end{eqnarray}
where $P_{pk,signal}$ is the peak power over the camera detector, ${P_{noise}}$ is the average power in the noise, and the camera dynamic range is denoted by $L$ ($L$= 12 for a 12-bit camera). In other words, while we continuously vary the power to change the PSNR, we discretize the inputs to the neural network to use the maximum range of the 12-bit camera.


\begin{wrapfigure}{r}{0.5\textwidth}
    \centering
    \includegraphics[width=0.4\textwidth]{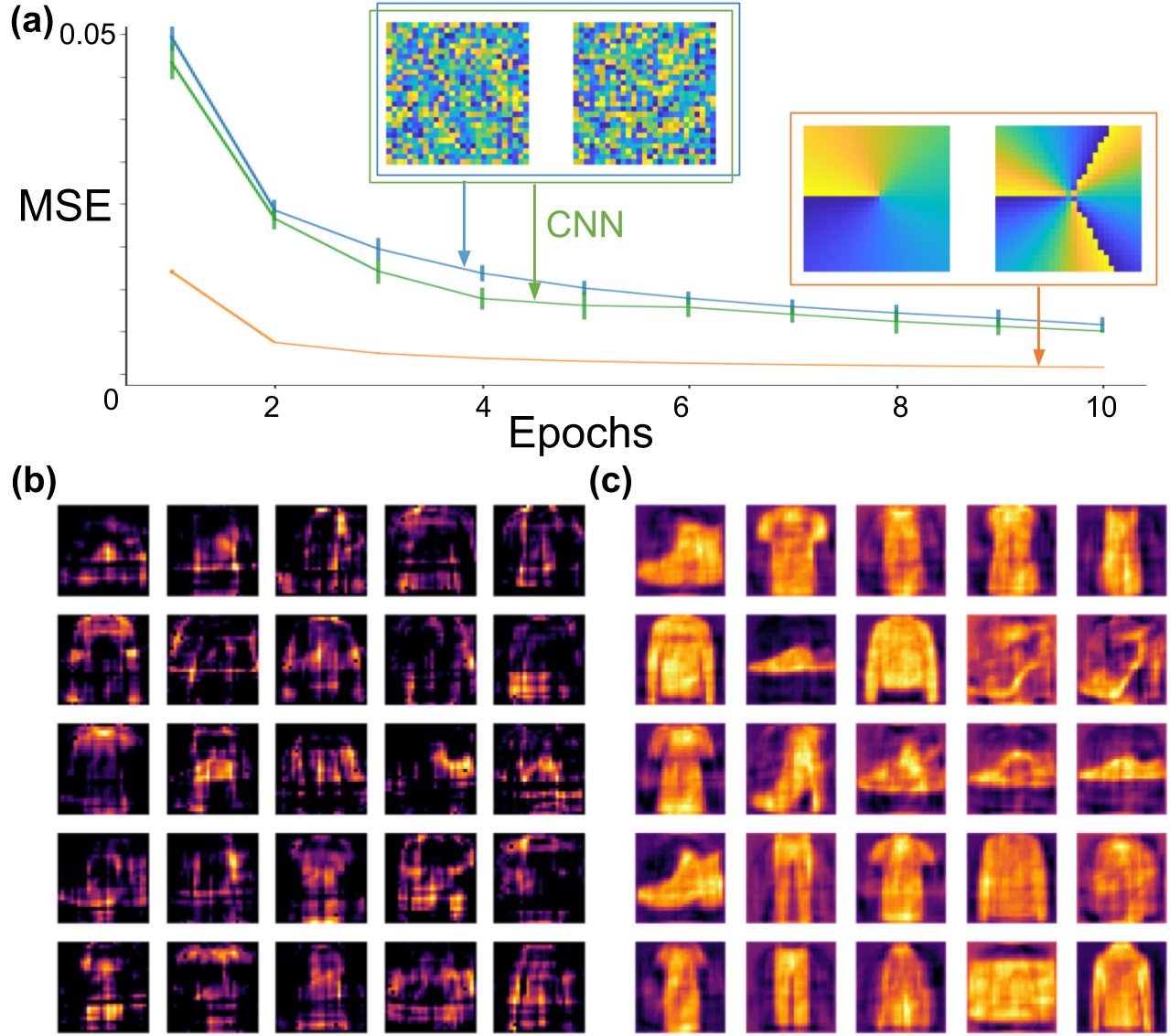}
    \caption{ (a) Small-brain convergence (MSE vs. epochs) training the reconstruction of MNIST fashion images with dual vortex $m=1, 3$ (red) and random-pattern (blue) spatial encoders. The random encoder converges at a similar rate with a convolutional neural network (green). The reconstructed images with (b) random-pattern and (c) vortex encoders with 2dB PSNR.}
    \label{Random_paterns}
\end{wrapfigure} 

Figure \ref{NoisyIm} illustrates the tradeoff between resolution and robustness to noise. With $m=0$ when the majority of the power is on-axis and there is no vortex, the signal intensity is most robust to noise but the neural network cannot solve the Phase Problem. With higher-$m$ (as well as larger $f\lambda$), the Fourier-plane pattern covers a larger area, so that the spectral features are sampled with better resolution. At the same time, when the area is larger, the effective PSNR decreases, resulting in a Fourier representation less robust to noise. 

Both with and without noise, we compare our results to a random spatial encoding pattern, where vortices are replaced with a diffuser, for example \cite{Antipa2017}. As with other spatial encoding schemes, the SSIM in reconstruction using the random phase patterns approaches the level of performance of vortex schemes in the no-noise scene when similar images are used to train and test the dataset. However, in order to achieve "near-vortex" performance, the random encoder requires more training time, which increases from 3 epochs to 8-10 epochs without noise. Furthermore, the situation changes drastically as we enter the noisy regime: while the vortex-Fourier encoding preserves image quality in the case of high noise, the random encoding scheme fails completely. The gains with accuracy also increase with the vortex encoding scheme, as shown in Figure \ref{Random_paterns} with 2dB PSNR. Thus, vortex encoding provides feature extraction for efficient reconstruction of the object $\mathbf{Y}$ to $\mathbf{X}$, which enables faster convergence of the neural network, as well as robust construction in the presence of noise.

\begin{wrapfigure}{r}{0.5\textwidth}
    \centering
    \includegraphics[width=0.4\textwidth]{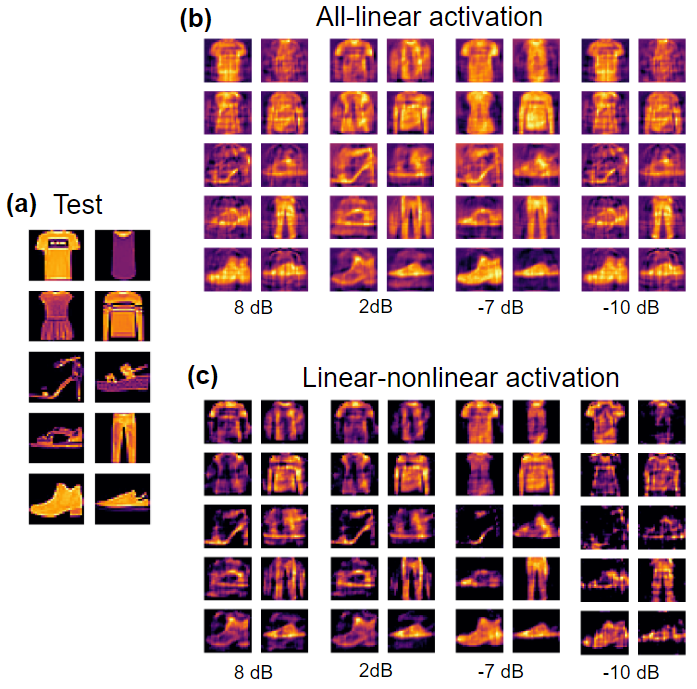}
    \caption{(a) Validation test images. Reconstruction results under PSNR-labeled noisy conditions that show the effect of (b) 'linear' and (c) 'nonlinear' activations for the final layer of the small-brain neural network. A vortex encoder with $m=1,3$ is used. }
    \label{Noise_rec}
\end{wrapfigure}

An interesting aspect of the inverse problem arises in the presence of noise when the neural network is trained with images without noise and also tested with images from the same dataset with noise. (Again, only two vortices are needed to produce the best reconstructed image.) The neural network learns an inverse mapping that minimizes MSE in the training set; however, different layer activations are vulnerable to different types of noise. This is illustrated in Fig. \ref{Noise_rec}. A linear activation leads to more ghosting and amplification of sensor shot noise, whereas a nonlinear activation is more vulnerable to the dark noise. While the nonlinear activation produces sharper images in the absence of noise, the linear activation is more robust and produces better images in the presence of more noise. With high noise, the nonlinear model's image quality is mixed.  We find that the most robust model uses linear activations in the hidden layer and nonlinear activations only for the last layer. The linear activations produce results that are more generalizable and transferable in the presence of noise. In contrast, while more accurate with low noise, nonlinear activations provide less of the ``inverse'' mapping, as seen by a bias in reconstruction for highlighting edges. 

A comparison of different random patterns as spatial encoders suggests that the use of completely stochastic patterns are not best suited for image reconstruction and the structured base sets, such as topological phases and vortices, are more effective for this task. This said, the work in \cite{Antipa2017} is often applied to incoherent light illumination for image processing. There is certainly space for further study of topological structures for light scattering that connect different spatial-encoding and compressive-sensing approaches with neural networks.

\section{Experiments and discussion}
In experiments, we capture Fourier-plane intensity patterns when the object illumination flux is too low to be imaged directly by the camera. In this way, the vortex imaging and small-brain deconvolution approach serves as a low-light-level camera, suitable for imaging through noisy environments. 

\begin{figure}
    \centering
    \includegraphics[width=0.9\textwidth]{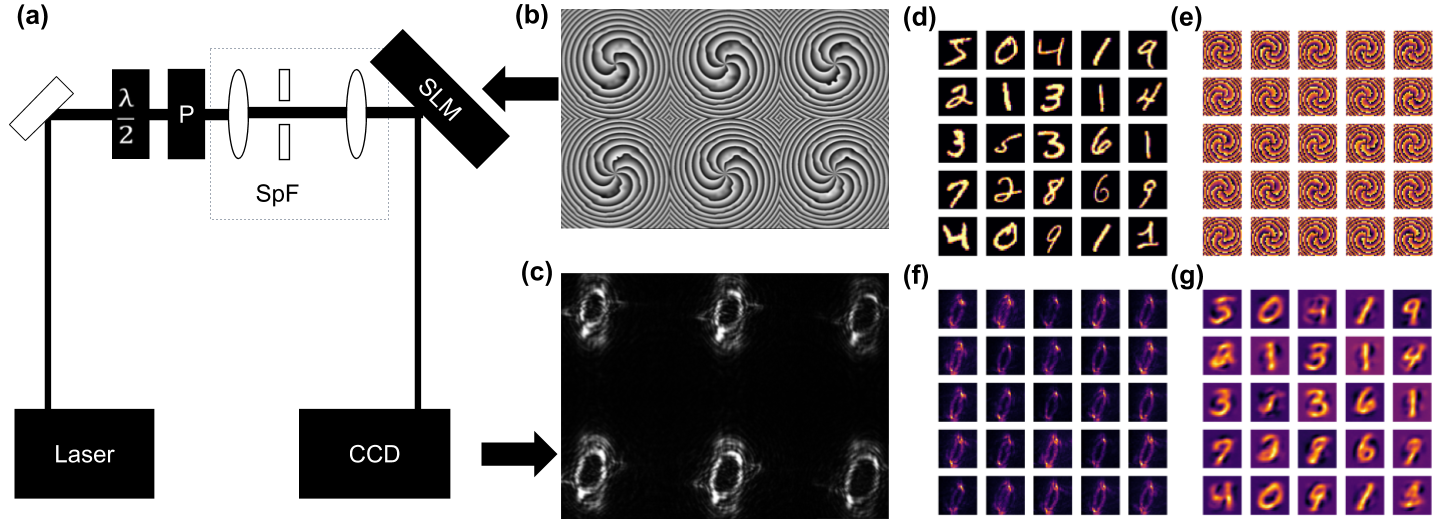}
    \caption{(a) Experimental setup. A laser ($\lambda = 532$ nm) is directed through the half-wave plate ($\lambda/2$),  polarizer (P), and spatial filter (SpF) to obtain a near-perfect Gaussian beam. The lens focuses the beam on the spatial light modulator liquid crystal matrix (SLM) that operates in reflection mode. (b) The SLM pattern that imprints an $m=4-9$ vortex, a quadratic radial phase, and MNIST handwritten digit from the PC. The reflected light is collected on the CCD camera and stored on the same PC. (c) Example CCD image for vortices $m=4-9$ (d) 25 test images, (e) the corresponding phase patterns, (f) resulting CCD images and (g) reconstructed images. }
    \label{Experiment}
\end{figure}

The experimental setup is shown in Fig. \ref{Experiment}(a). Coherent polarized light from the laser source (diameter: 1.1mm) is reflected from the two mirrors, passes through the half-wave plate and a second polarizer, which rotates to vary the power. The laser is a 500-fs pulsed Nd:YAG Fianium at its second harmonic, $\lambda = 532$ nm. A spatial filter eliminates parasitic modes. The transmitted, linearly-polarized $TEM_{0,0}$ beam is collimated (diameter: 1.5cm) onto the spatial light modulator (SLM)(Hammamatsu LCOS-SLM) at an incident angle of 30$^\circ$. The SLM has 800$\times$600 pixels. 

When recording experimental data, the phase patterns are saved in files at the start. An example phase pattern is shown in Fig. \ref{Experiment}(b). Each pattern contains 6 lenslet images with vorticial ($m=4-9$) and quadratic radial phases ($f/\lambda$ = 0.1), as well as the imprint of an MNIST handwritten digit. An automated computer program sends the phase pattern from the saved file to the SLM, while a second program records the reflected light and grabs the CCD camera image. The CCD camera (Thorlabs-DCU223M) is approximately 20 cm from the SLM and has 4.65$\mu$m$\times$4.65-$\mu$m pixel area, 8-bit dynamic range, and 1024$\times$768 pixel resolution. The oblique reflections of the SLM phase pattern result in vertically-elongated images on the CCD.

We achieve impressive results with this 8-bit CCD camera. Imaging is demonstrated here with average light fluxes of 30 nJ/cm$^2$ (intensities of 10$\mu$W/cm$^2$, exposure times of 2.8 ms) over 6 vortex masks at the SLM. A ``reference'' CCD image that is taken with higher intensity is shown in Fig. \ref{Experiment}(c). This Fourier-plane image is used to center the cropped data since with such low illumination fluxes, we use only 5-10\% of the dynamic range of the 8-bit camera and the images are virtually all dark. For comparison, the signal flux at the SLM would be below the level of the camera read noise, even with minute-long camera exposure intervals. 

With our automated matlab program, we record approximately 5 images per second for the training and testing of our algorithm. The primary limitation with speed is the camera retrieval time, which is much longer than the camera shutter time. The delay time between image uploads and the presence of SLM phase patterns is 50 ms. To achieve experimental results analogous to the simulated efforts, the CCD image subsets are cropped and downsized with 'inter-area' interpolation to 28x28 pixels for each $\mathbf{Y}$. A square area is cropped even though the CCD images are elongated.  The validation/test, reconstructed, and SLM/CCD images taken with 10-$\mu$W power and 2.8-ms exposure times are shown in Figs. \ref{Experiment}(d-g). All images are normalized to the color-range since without this colorbar normalization, the vortex Fourier patterns would not be visible. Even though there are 6 vortex lenslets in the CCD images, we use only one vortex $m=4$ and achieve an SSIM of 0.688 with all-linear and linear-nonlinear activations. We use only 4500 (500) images from the dataset for training (testing) even though there are 60,000 images in the MNIST dataset. In fact, the training of the algorithm converges to similar values of accuracy, without overfitting, with only 2000 training images. 

While we could integrate more images from the dataset or more patterns from different-$m$ vortices to train the small-brain neural network, doing so would not improve the accuracy of the reconstruction algorithm, which at this low-light level is limited by the camera characteristics. Similarly, we do not see significant improvement by varying the activations of the neural network layers. In other words, the primary limitation is the sensitivity of our camera. In this low-light level range, we are using only 4-bits of the camera (unlike in simulations where we assumed the full dynamic range of a 12-bit camera). While the robustness to overfitting in the training algorithm is impressive i.e., we need only 2000 images to train the algorithm, the training time increases by a factor of 4 compared to simulations. Longer training times may arise when the MSE gradients are shallow and when the dynamic range of the neural-network inputs are limited.

Once trained, we still reconstruct MNIST Fashion images from a vortex Fourier representation at a rate of several thousand frames per second on a 15W central processing unit, two orders of magnitude faster than convolutional neural net schemes. Our approach to and understanding of image reconstruction with the topological Fourier encoding and small-brain neural network is new. Future work should include the effects of different types of noise, variations between training and test sets, topological phase and compression, and camera characteristics on the robustness of the neural-network reconstruction. 

\section{Conclusion}

We present a vortex-Fourier encoding approach to pre-processing data prior to a neural network. The phase singularity is imprinted on the object prior to the lens Fourier transform. As a result of the spiral phase, we observe edge-enhanced, compressed, and phase-preserved representations. While many inverse problems are solved iteratively or involve CNNs, we show that it is possible to solve inverse problems and obviate the need for CNNs with Fourier representations of vortex-encoded objects. Spatial encoding with a topological phase results in efficient feature extraction of the Fourier pattern and accelerates learning for the inverse reconstruction of the object when random encoders do not. Our approach, using shallow, dense neural networks or ``small-brain'' machine learning offers a strategy for accurate, robust, and rapid camera-like imaging in low-light or noisy environments. 

The specific spatial encoding by a vortex provides mixed edge-enhanced real and imaginary components, so that with 2 vortices, the neural network is transferable and generalizable. We aim to unbox the black box of machine learning by solving the generalized problem of inverse construction with categorically similar and dissimilar images and different MNIST datasets. We also show that there are different levels of robustness to different types of noise/deviation from training sets with different layer activations. Nevertheless, our small-brain machine-learning algorithm reduces the computational overhead with training and also reduces computational complexity in reconstruction, resulting in images being less vulnerable to adversarial attacks. 

To summarize, the optical preprocessing approach demonstrated here with a topological phase mask and lens is:
\begin{itemize}
\item \textbf{Robust to noise.} Signals effectively achieve 200-2000X higher PSNRs. We successfully capture and deconvolve objects illuminated with $10 \mu$W/$cm^2$ average intensities with millisecond shutter times using 4 bits of an 8-bit CCD camera.
\item \textbf{Single shot.} Reconstruction is possible with a single image containing two vortices or two orthogonal topological phases.
\item \textbf{Low-latency and fast.} Our approach has potential for real-time processing and video-camera streaming. With Fashion MNIST images, we process several thousand frames per second with low-power hardware (10-20W).
\item \textbf{Computationally efficient.} While other methods currently take multiple encoded images or use iterative schemes, we achieve near-to-ideal reconstruction with two ptychographic images.
\item \textbf{Extremely low-power computation.} The technique uses explicitly simple neural nets (no deep learning) where pre-processing is completed with parallel optical propagation.
\item \textbf{Compact with memory.} The vortex Fourier transform provides a compressed representation that can further be leveraged to minimize the number of pixels that carry data forward.
\item \textbf{Flexible with a digital re-adjustable stage.} There is a tradeoff between resolution, robustness, and sampling that we control with the choice of vortex charge $m$ and focal length $f$.
\end{itemize}

If this approach is successful at reconstructing real objects with depth of field, there are numerous applications that involve imaging in low signal conditions, such as security-related systems where illumination is minimal, driver-assist systems, microscopy of delicate photosensitive biological samples, among others. Given the low power requirements and high frame-rate reconstruction speeds, our scheme is expected to be useful for satellite or unmanned operations. One envisions that the vortex-Fourier encoding scheme may be efficient at collecting radiation in pulsed, spectroscopic, laser applications. Further research may exploit topological features for achieving greater depth of field \cite{Greengard2006, Pavani2008} and extend our knowledge of vortex Fourier ptychography to leverage information in light polarization, dispersion, and spatio-temporal coherence.

---------------------------------------




\section{acknowledgements}
The authors acknowledge editing support from Ben Stewart $\langle$linkedin:benjamin-w-stewart$\rangle$.

\section{Funding}
 The authors gratefully acknowledge DARPA YFA Grant $\#$D19AP00036.
\section{disclosures}
 The authors declare no conflicts of interest.

\end{document}